\documentclass[aps,prd,preprintnumbers,nofootinbib,twocolumn]{revtex4}
\usepackage{bm}
\usepackage{latexsym}
\usepackage{dcolumn}
\usepackage{amsmath,amsfonts,amssymb}
\usepackage{graphicx,epsfig}
\usepackage{psfrag}
\usepackage{color}
\usepackage{soul}
\usepackage{ulem}
\usepackage{amsthm}

\begin{document}

\title{{\LARGE Bose-Einstein condensate in cosmology}}

\author{Saurya Das $^1$%${}^\star$ (saurya.das@uleth.ca)
}
\email{saurya.das@uleth.ca}

\author{R. K. Bhaduri $^2$}
\email{bhaduri@physics.mcmaster.ca}

\affiliation{
\vspace{0.3cm}
$^1$
Theoretical Physics Group and Quantum Alberta,
Department of Physics and Astronomy, University of Lethbridge, \\
4401 University Drive, Lethbridge, Alberta, T1K 3M4, Canada  \\
{}\\
$^{2}$
Department of Physics and Astronomy,
McMaster University, Hamilton,
Ontario, Canada L8S 4M1}

%\date{}

%\sf

\begin{abstract}
Applying the seminal work of Bose in 1924 on what was later known as
 Bose-Einstein statistics,
Einstein predicted in 1925 that at sufficiently low temperatures,
a macroscopic fraction of constituents of a gas of bosons will drop down
to the lowest available energy state,
forming a `giant molecule' or a Bose-Einstein condensate (BEC),
described by a `macroscopic wavefunction'.
In this article we show that when the BEC of ultralight bosons extends over cosmological length scales,
%its maximally allowed size,
%namely that of the visible universe,
it can potentially explain the origins of both dark matter and dark energy.
%We describe pros and cons of our model and
We speculate on the nature
of these bosons.
{}\\
{}\\
{\bf Invited review written
for the special issue of
{\it Physics News} on the occasion of
125${}^{th}$ birthday anniversary of S. N. Bose.
}
\end{abstract}

\maketitle

\section{Introduction}

Two of the enduring mysteries in cosmology are dark matter (DM) and dark energy (DE).
Whereas DM holds the rotating galaxies together, DE makes the expanding universe
accelerate. In accounting for the distribution of mass/energy in the universe,
visible hadronic matter and radiation contribute only about five percent, DM about twenty five
percent, while the rest, a whopping seventy percent, comes from DE
\footnote{For alternative theories to DM and DE which also try to explain cosmological observations,
see
\cite{mond1,mond2,altdm,altde}.}.
To start with, the constituents of
DE are not known, despite viable candidates such as the cosmological constant and a dynamical scalar field \cite{dereview1,dereview2}.
The constituents of DM are not known either.
There has been many studies invoking weakly interacting
massive particles (WIMPs) that may form cold DM.
Not only does it have shortcomings in reproducing
the DM density profiles within a galaxy, no such particle has been experimentally found.
Other DM candidates include solitons, massive compact (halo) objects,
primordial black holes, gravitons etc. They have similar shortcomings \cite{dmreview1,dmreview2}.

Given that DM is all-pervading, cold and dark, and clumped
near galaxies, we pose the following question:
can it be a giant BEC, of cosmological length scales?
Following the paper by Bose which laid the
foundations of Bose-Einstein statistics \cite{bose},
Einstein predicted that a gas of bosons
will form a BEC at sufficiently low temperatures
\cite{einstein}.
Here we show that following an earlier proposal by the current authors \cite{db1}, \\
(i) a sea of weakly interacting light bosons can
form a BEC, preserving large scale homogeneity and isotropy,
and be a viable DM candidate,
as long as the mass of each constituent does
not exceed a few $eV/c^2$
\footnote{$1$ kg = $5.6\times 10^{35}~eV/c^2$, where $c=3\times 10^8~m/s$ is the speed
of light in vacuum.}, and \\
(ii)
%for the above
%constituent mass,
%of $10^{-32}~eV/c^2$,
the quantum potential associated with the above BEC
can also explain DE.

Note that in a BEC, the bosons are in the
lowest energy state that is nearly at zero energy,
even if outside the condensate the
bosons are highly relativistic due to its low mass.
We should point out that BEC model for DM has
been studied by many authors in the past.
For example, one can consider a scalar field
dark matter (SFDM) that invokes spin zero ultralight bosons whose
Compton wave length spans cosmic
distances \cite{Hu2000,Lopez,Bohua}.
For previous studies of superfluids and BEC in cosmology see
\cite{sudarshan,khlo1,khlo2,morikawa,moffat,wang,boehmer,sikivie,chavanis,dvali,houri,kain,suarez,ebadi,laszlo1,bettoni,gielen,schive,davidson,casadio1,casadio2}.
But as mentioned earlier, the
novel aspect of our model is that it also provides us with a viable source of DE.
In this semi-technical article, we present the bulk of our work
in an elementary fashion using Newtonian dynamics, after staging the
backdrop of the cosmological model.
%We present a rigorous relativistic derivation in the Appendix.
%
%In Newtonian cosmology, the observer is at the center of the universe,
%whereas in General Relativity (GR) all points in space are equivalent
%\cite {Arnab}.
A more rigorous derivation of essentially the same results
using general relativity will be presented in the Appendix.

\section{Cosmology}

To put the problem in perspective, we quickly review an elementary description  of
cosmology. On scales of about $300$ Mpc or higher
\footnote{$1$ Mpc $= 3.3 \times 10^6$ light years.},
the distribution of matter in the
visible universe is found to be homogeneous and isotropic to a very high degree,
about $1$ part in $10^5$.
One can assume this matter to be a perfect fluid,
described by the equation of state
$p=w\rho c^2$, where $p$ and $\rho$ are its pressure
and density respectively. The constant $w$ characterizes the fluid, e.g. $w=0$ for baryonic matter (i.e. galaxies etc) and dark matter
(both being non-relativistic for which
$\langle v^2 \rangle/c^2 \approx 0$),
$w=1/3$ for relativistic matter (such as
cosmic microwave photons and neutrinos) and
$w=-1$ for a substance which exerts negative pressure and as we shall see later, makes the universe accelerate (e.g. dark energy, cosmological constant and a scalar field).
Moreover, by observing the red shift in deep space, Hubble found that distant stars and
galaxies were receding from each other, with the speed of recession
proportional to their distance. This meant that (along the line of
sight) $\dot{r}=H r$, where H is independent of $r$, but can be a function of
time $t$. We can therefore write $r=r_0~a(t)$ where $r_0$ has the dimension of length,
and $a(t)$ is the so-called scale factor, which characterizes all distance scales in
the universe including distance between galaxies at any given point of time.
Therefore the Hubble ``constant''
$H \equiv \dot{a}/a$ has the dimension of inverse time.
Clearly, $H>0$ for an expanding universe, and vice-versa.
The most general spacetime metric consistent with the
observed large-scale homogeneity and isotropy is of the following form,
also known as the Friedmann-Robertson-Walker metric, and
written in spherical polar coordinates:
\begin{eqnarray}
ds^2= c^2 dt^2-a^2(t)\left( \frac{dr^2}{1-\kappa r^2} + r^2 (d\theta^2 + \sin^2\theta d\phi^2) \right)~,
\label{metric1}
\end{eqnarray}
where $\kappa=0,1,-1$ for spatially flat, positive curvature (a three-sphere) and negative
curvature (a three-dimensional saddle) respectively \cite{weinberg}.
Further, it has also been
determined from the Supernova and other observations
that the spatial part of the universe at any given epoch in time is flat, i.e.
$\kappa=0$ in Eq.(\ref{metric1}) \cite{perlmutter,riess}.
This implies that the universe is and has always been of infinite extent and without a boundary.
$\kappa=-1$ also means an infinite universe, while
$\kappa=1$ signifies a closed and finite-sized universe with no boundaries.
As we shall show in the next section, $\kappa=0$ implies that the total effective density $\rho=\rho_{crit}$,
where $\rho_{crit} \equiv 3H^2/8\pi G$ is known as the critical density.
In the present epoch, substituing
$H=H_0$ the current Hubble parameter, one gets
$\rho_{crit} \approx 10^{-26}~kg/m^3.$
%
%\begin{eqnarray}
%ds^2 && =c^2 dt^2-a^2(t)(dx^2+dy^2+dz^2)~ \\
%%
%& = c^2 dt^2-a^2(t)( dr^2 + r^2 (d\theta^2 + \sin^2\theta d\phi^2) )~,
%\end{eqnarray}
%

In the adjoining Fig.1, we show how the time-slices can be made at various times to
show the expanding space as time advances.
If light of wavelength $\lambda_e$ is emitted at time $t$ and observed at a
later time by a comoving observer
(i.e. an observer in whose frame the universe appears homogeneous and isotropic),
it is found to have a larger wave length $\lambda_0$,
since the wave length stretches out with the scale factor $a(t)$.
We can then write the ratio
\begin{equation}
{\lambda_0/\lambda_e}=a_0/a_e \equiv 1+z~,
\end{equation}
where $z>0$ for red shift.
One can make an estimate of the age of the universe by assuming that it has
been expanding (approximately) linearly with time at its present rate from the beginning.
Under this approximation, and ignoring quantum effects in the very early universe,
one can define the beginning or
the moment of the big bang when the scale factor was zero, and the
density of the universe was infinite.
%all of the universe was
%concentrated at a point.
%Fig.2 shows a linear increase of the scale factor with time.
%Interpolating backwards, it shows that the big bang at time
%$H_0^{-1}$.
%With the currently accepted value of $H_0=70~(km/s)/Mpc$, this is about
%$3\times 10^{17}~s$, or $10$ billion years, which is quite close to the more sophisticated
%estimate of $14$ billion years, using general relativity and detailed assumptions of the matter
%content of the universe.
%
One can then define a Hubble length $L_0=c/H_0$, where the
subscript zero signifies the present values.
The size of the visible universe is
identified as $L_0$. Using the present value of the Hubble constant of about $70$ km/s/Mpc,
we get the age of the universe to be $13.8 \times 10^9$ years, and its size $L_0$ to be
$1.3 \times 10^{26}$ meters.

%\vs{0.9cm}
\begin{center}
\begin{figure}
\includegraphics[scale=0.15,angle=0]{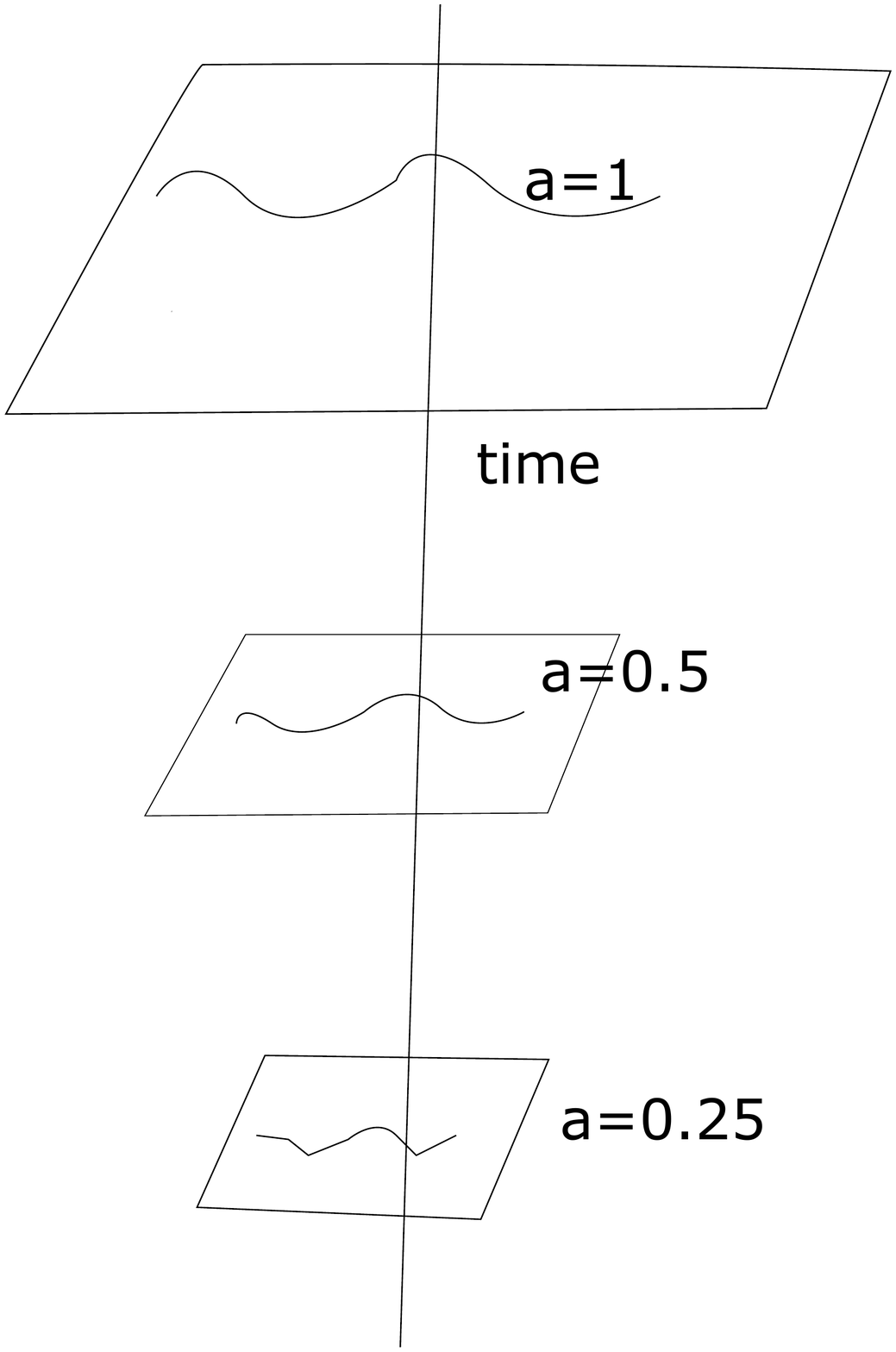}
\caption{Expanding scale factor and wavelengths with time in a spatialy flat universe}.
\label{solution}
\end{figure}
\end{center}

\subsection{Newtonian Cosmology}

We now describe the dynamics of the expanding universe in terms of a simplified
Newtonian model, which as we shall see captures the essential ingredients.
Consider a sphere of mass $M$ and radius $r$ with uniform density $\rho$
anywhere in a homogeneous and isotropic universe, and
a point mass $m$ on its surface.
Using Newton's second law, we get
\begin{equation}
m \ddot{r}=-G~M~m/r^2.
\end{equation}
Using the relation that $M=\frac{4 \pi}{3} \rho r^3$ and $r=r_0~a(t)$, we get
\begin{equation}
\frac{\ddot{a}}{a}= -\frac{4 \pi G \rho}{3}~.
\label{fried1}
\end{equation}
In the cosmic fluid model, it is the stress energy tensor that is important
(or more accurately its trace, or sum of its diagonal elements),
rather than the mass density $\rho$.
In the comoving frame, only the diagonal
elements $T_{\mu\mu}/c^2=(\rho,p/c^2,p/c^2,p/c^2)$ are nonzero,
where $p$ refers to pressure. We then obtain
\begin{equation}
\frac{\ddot{a}}{a}=-\frac{4 \pi G}{3} \left(\rho + \frac{3p}{c^2} \right)
\label{cos5}
\end{equation}
The above equation shows that there is deceleration due to gravity alone.
This is to be expected since gravity is an attractive force. Measurements of red
shifts from very distant type 1a supernovae explosions show however
that $\ddot{a}>0$, i.e. the universe is not just expanding,
but also accelerating
\cite{perlmutter,riess}.
This could happen if we introduce a repulsive potential that overcomes the
attractive gravitational force at very large distances.
This could arise
from Einstein's equation with a cosmological constant $\Lambda$ \cite{Rindler}.
%\begin{equation}
%(R_{\mu\nu}-1/2 g_{\mu\nu} R)+\Lambda g_{\mu\nu}=-8\pi G/c^4 T_{\mu\nu}~.
%\end{equation}
The starting equation of motion is then given by
\begin{equation}
\ddot{r}= -  \frac{GM}{r^2} + \frac{\Lambda c^2~r}{3}~.
\end{equation}
The corresponding potential is
(this also follows from the identification of the potential
$\frac{V(r)}{m}
= - |g_{00}-1|c^2/2$, where one reads off
$g_{00}$ from the metric of de Sitter space):
\begin{equation}
\frac{V(r)}{m}
= -\frac{GM}{r} - \frac{\Lambda c^2r^2}{6}~.
\label{cos1}
\end{equation}
Then including the cosmological term, Eq.(\ref{cos5}) gets modified to
\begin{equation}
\frac{\ddot{a}}{a}=-\frac{4\pi G}{3} \left(\rho+ \frac{3p}{c^2} \right)+\frac{1}{3}\Lambda c^2~.
\label{cos6}
\end{equation}
As one can see from the above, a sufficiently large
and positive $\Lambda$ would result in $\ddot a>0$,
i.e. an accelerating universe, just as observed.

Further, using the first law of thermodynamics for isentropic expansion,
namely $dU+pdv=TdS=0$ (where $U$ is the internal energy, $v \propto a^3$ the volume,
$T$ the temperature and $S$ the entropy) and the equation of state $U=\rho v$,
one gets $\dot \rho + 3(p+\rho) \dot a/a =0$. Using Eq.(\ref{cos6}) to
eliminate $p$, one obtains
\begin{equation}
H^2=\left( \frac{\dot{a}}{a} \right)^2= \frac{8\pi G \rho}{3} + \frac{\Lambda c^2}{3}
- \frac{\kappa c^2}{a^2} ~,
\label{frw2}
\end{equation}
where the integration constant
$\kappa$ is $+1,0,-1$ for spatial curvature to be positive, flat, or
negative.
The above equation can be conveniently written as
\begin{eqnarray}
\Omega(t) - 1 = \frac{\kappa}{(aH)^2}~,
\end{eqnarray}
where $\Omega(t) \equiv \rho(t)/\rho_{crit}$, $\rho(t)$ being the total effective density of matter, radiation
and the cosmological constant at any epoch, and as mentioned earlier $\rho_c \equiv 3H^2/8\pi G$.
Thus, $\rho=\rho_b+\rho_{DM}+\rho_{\Lambda}$, where $\rho_{\Lambda}=
\Lambda c^2/(8\pi G)$. Here $\rho_b$ is for visible baryonic matter or
radiation, $\rho_{DM}$ is for DM, and $\rho_{\Lambda}$ is for DE \cite{weinberg}.
In the simplest situation (with no intervening medium), knowing the absolute luminosity of the source,
and measuring its apparent luminosity should allow one to figure out how far it is from the observer.
There is an added complication however,
because the intervening medium has different components of densities
$\rho_M, \rho_b, \rho_\Lambda$ evolving with different $z$-dependences,
as light reaches the observer from the source.
One has to integrate the square rooted Eq.(\ref{frw2}) to find out the contributions of the densities to get the
``luminosity distance", or the effective distance from the source taking into account all intervening matter.
When this distance is plotted against the red shift $z$ for `standard candles' such as Supernovae 1a
(whose intrinsic luminosities are known to a high degree of accuracy),
the best fit is found for $\rho_\Lambda/\rho_{crit}=0.70, \rho_{DM}/\rho_{crit}=0.25, \rho_b/\rho_{crit}=0.05$.
Further, since high $z$ measurements show $\kappa=0$, and $\Lambda >0$,
one has $\rho=\rho_{crit}$ exactly {\it and for all} epochs! This is another mystery in itself, which
theories such as inflation attempt to resolve \cite{inflation}. We will not address this here.
%
%Moreover, from luminosity observations of `standard candles' such as Supernovae 1a,
%one estimates the following fractions
%$\rho_{b}/\rho_c = 0.05$, $\rho_{DM}/\rho_c = 0.25$ and $\rho_\Lambda/\rho_c = 0.70$.
%
One can now write  Eq.(\ref{frw2}) as
\begin{equation}
H^2=\frac{8\pi G \rho_{crit}}{3}
\label{frw3}
\end{equation}
From the above, one gets the cosmological constant
$\Lambda= 8\pi G \rho_\Lambda/c^2  =10^{-52}~(meter)^{-2}.$
Before we link this tiny constant to quantum corrections to the classical GR
equations, we elaborate on our model of BEC in cosmology.

\section{BEC in Cosmology}

We conjecture that there are ultra light bosons that span our universe, their
Compton wavelength is comparable to the size of the visible universe, and that
a BEC of these bosons constitute DM.
We will discuss about the nature of these bosons later.
First consider
the case where the bosons are their own antiparticles, e.g. for bosons with no charge.
%COPY FROM OUR PUBLISHED PAPER (Class. Quantum.Grav 32 (2015) 105003 , page 2
%Top paragraph  starting
%``To compute the critical temperature.......and behave as CDM''.

To check if these bosons can indeed form a BEC,
we first compute their critical temperature $T_c$ and compare it with the ambient
temperature of the universe,
that of the all pervading cosmic microwave background radiation (CMBR) at any epoch
$T(a)=(2.7/a)$ K (we assume $a=1$ in the current epoch).
A bath of bosons forms a BEC if its temperature falls below the critical temperature
\footnote{The BEC critical temperature should not be confused with the critical density in cosmology, although both are used in this article.}.
Therefore if $T(a)<T_c$ at some epoch, then a BEC of these bosons would form at that epoch,
and if the inequality holds for future epochs, then the BEC will continue to exist.
The boson must have a mass, however small.
In a BEC, the
average inter-particle distance $(N/v)^{-1/3}$ (where $N=$ total number of bosons in volume $v$)
is comparable or smaller than the thermal de Broglie wavelength $hc/(k_B T)$, such that
quantum effects start to dominate.
Identifying this temperature of a bosonic gas to the critical temperature $T_c$,
we get $k_B T_c \simeq h c (N/v)^{1/3}$.
A more careful calculation for ultra-relativistic
weakly interacting bosons with a tiny mass gives
\cite{brack,grether,fujita}
\begin{eqnarray}
T_c = \frac{\hbar c}{k_B} \left( \frac{N \pi^2}{v \eta \zeta(3)} \right)^{1/3}~.
%= \frac{\hbar c}{k_B} \le( \frac{N \pi^2}{V_0 \eta \zeta(3)} \ri)^{1/3} \frac{1}{a}
%\equiv \frac{T_{c0}}{a}
\label{tc1}
\end{eqnarray}
%
%\vs{0.9cm}
\begin{center}
\begin{figure}
\includegraphics[scale=0.45,angle=0]{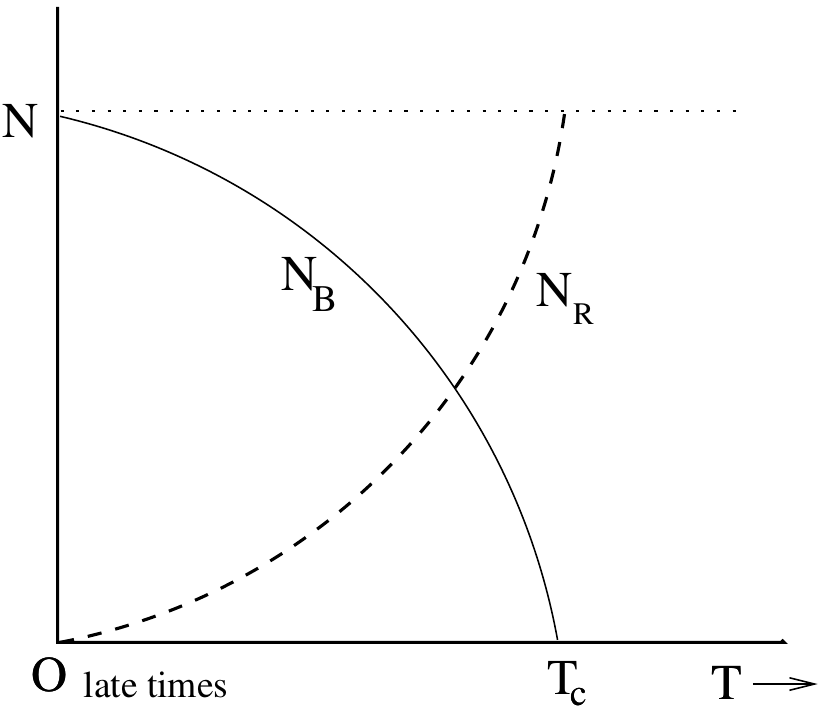}
\caption{$N_B$ and $N_R$ vs. $T$.}.
\label{solution}
\end{figure}
\end{center}
%
%``''
In the above
$N=N_B + N_R$, $N_B$ being the number of bosons in the BEC, and $N_R$ outside it,
both consisting of bosons of small mass as discussed earlier,
$\eta$ is the polarization factor
(which is $1$ for scalars and $5$ for massive gravitons) and
$\zeta(3) \approx 1.2$.
Also
%$a$ is the cosmological scale factor,
$L=L_0 a$ is the Hubble radius and
$v=L^3=L_0^3 a^3 \equiv v_0 a^3$.
(subscript $0$ here and in subsequent expressions denote current epoch,
when as we mentioned earlier $a=1$). % $L_0=1.4 \times 10^{26}~metre$).
Note that for boson temperature $T<T_c$, a BEC will necessarily form
%even when there are interactions
\cite{bhaduricjp}.
As stated before, identifying the BEC of bosons all in their ground states with zero momenta and only
rest energies with DM in any epoch (i.e. $N \simeq N_B$), one gets
$N_B/v= \rho_{DM}/m =0.25 \rho_{crit}/ma^3$
% $\simeq 0.7 \times 10^{42}~a^{-3}~(metre)^{-3} \gg N_R~,$
%
Then from Eq.(\ref{tc1}) and the relation $T(a)<T_c$ one obtains
%
%\begin{eqnarray}
%T_c = \frac{6 \times 10^{-12}}{m^{1/3}~ a}~K~,
%\label{tc2}
%\end{eqnarray}
%
%
\begin{eqnarray}
T_c = \frac{4.9}{m^{1/3}~ a}~K~,
\label{tc2}
\end{eqnarray}
($m$ in $eV/c^2$ in the above).
Note that both $T_c$ in Eq.(\ref{tc2}) and the CMBR temperature $T(a)$ scale as
$1/a$, the inverse of the scale factor.
Therefore $T(a)<T_c$ gives $m<6~eV/c^2$
(or about $m<10^{-35}$~kg),
independent of $a$.
In other words, for a gas of bosons which weigh a few electron volts
or less, its critical temperature will always exceed the temperature of the
universe, it will form a BEC at very early epochs and is a viable DM candidate.
Having little or no
momentum in the BEC, they will behave as cold dark matter (CDM).
Furthermore, $m=10^{-32}~eV/c^2$, the boson
mass required for its quantum potential to account for the observed cosmological constant is well
within the above bound!
Fig.2 shows the proportion of bosons in the BEC ($N_B$) and that outside the BEC ($N_R$) at any temperature $0 \leq  T \leq T_c$. As the universe temperature drops (towards the left of the diagram), more and more bosons drop to the BEC and in the far future, almost all are in the BEC state.

We would like to point out
that our results are robust and do not change much, even if we consider the case when the bosons and their antiparticles (antibosons) are identifiable
(this can happen even for neutral bosons if they have a hidden quantum number). In this case, computing the critical temperature
\cite{grether,Haber}
and comparing with $T(a)$, one obtains
$m<2.4\times 10^{-2}~eV/c^2$ for BEC to form in the present epoch and a larger upper bound at earlier epochs. The corresponding number for
non-relativistic bosons obtained from its critical temperature \cite{landau}
is $m<3.8\times 10^{-3}~eV/c^2$.
In either case, the mass required for DE is included in the above bounds. We note however that the $a$-dependences of $T_c$ are different in these cases, for which a BEC made up of these constituents would fragment in the far future.

\section{DE and Quantum Potential}

We noted in Eq.(\ref{cos1}) that the repulsive term arose from the cosmological
constant $\Lambda$. We shall now show that in our model this term, responsible for
DE comes from the simplest quantum correction in the trajectory of the bosons in
the BEC. A key point is that every boson in the BEC has the same wave function.
Furthermore, once inside the BEC,
the bosons, however light, are slow and non-relativistic.
This makes it possible to use the single particle wave function $\Psi$ that obeys the
one-body Schr\"odinger equation
(our arriving at the same results using the relativistic wave equation in the Appendix confirms this)
\begin{equation}
-\frac{\hbar^2}{2m} \nabla^2 \Psi + V \Psi = i\hbar\frac{\partial \Psi}{\partial t}.
\label{cos2}
\end{equation}
Let us write $\Psi$ as
\begin{equation}
\Psi (\vec r,t)={\cal R} \exp (iS/\hbar)~,
\end{equation}
where ${\cal R} (\vec r,t)$ and $S (\vec r,t)$ are real functions.
We then obtain
\begin{equation}
-\frac{\partial S}{\partial t}=\frac{(\nabla S)^2}{2m}+ V(r)
-\frac{\hbar^2}{2m}
\frac{\nabla^2 {\cal R}}{{\cal R}}~,
\label{cos3}
\end{equation}
and the continuity equation
\begin{equation}
\frac{\partial \rho}{\partial t}+\nabla.(\rho \frac{\nabla S}{m})=0~,
\end{equation}
with $\rho={\cal R}^2$.
If the last term in Eq.(\ref{cos3})
\begin{equation}
V_Q=-\frac{\hbar^2}{2 m}
\frac{\nabla^2 {\cal R}}{{\cal R}}
\label{cos4}
\end{equation}
is omitted, we are left with the classical Hamilton-Jacobi equation, giving rise to
the standard Newtonian trajectories for the constituent particles.
Note that $v (\vec r,t)\equiv (\hbar/m)\vec\nabla S$ can be identified with the `velocity field',
tangent to the streamline at any point within the BEC.
Inclusion of the `quantum potential' term (\ref{cos4})
on the other hand gives rise to `quantal trajectories' or Bohmian trajectories
for the constituents (named after its discoverer David Bohm).
These are ${\cal O}(\hbar^2)$ corrections to the
Newtonian trajectories which become important at short distances
(when quantum effects are expected to dominate) and indeed reproduce
the results of all quantum experiments, such as the interference patterns
in a double slit experiment. This interpretation of quantum mechanics,
albeit not the most popular one, is perfectly equivalent to the standard ones,
and offers a novel and useful picture in many situations,
such as the current one \cite{bohm,holland}.

Next, it was shown by one of the current authors
that inclusion of the quantum potential $V_Q$
resulted in the `Quantum Raychaudhuri equation'
\begin{eqnarray}
\frac{d\theta}{dt} = -\frac{1}{3} \theta^2 + \frac{\hbar^2}{2m^2} \nabla^2
\left( \frac{1}{{\cal R}}~\nabla^2 {\cal R} \right)~,
\label{qre1}
\end{eqnarray}
where the expansion $\theta$ measured the fractional change of
cross sectional area of a pencil of trajectories, as one moves along them.
In the absence of quantum corrections,
this equation with its negative definite term in the
right hand side predicts
$\theta \rightarrow -\infty$ for some finite time $t=t_0$. In other words,
particle trajectories in the presence of gravity converge in a finite time.
While this is expected since gravity is universally attractive, the result has
profound consequences for the nature of spacetime and the theory of relativity itself, as
we shall briefly touch upon in the Appendix.
Since the expansion or contraction of the universe
can also be described in terms of divergence or convergence of particle trajectories (those of the perfect fluid), the Friedmann equations of cosmology can also be derived from the Raychaudhuri equation
via the identification
$\theta = 3\dot a/a$ \cite{akrbook}. From the
quantum Raychaudhuri equation above on the other hand, one obtains
the quantum corrected Friedmann equation \cite{das1,das2}
\begin{equation}
\frac{\ddot{a}}{a}=-\frac{4\pi G}{3}(\rho+\frac{3p}{c^2}) + \frac{\hbar^2}{6 m^2}~
\nabla^2
\left(
\frac{\nabla^2 {\cal R}}{{\cal R}}
\right)~.
\end{equation}
Comparing with Eq.(\ref{cos6}), we write
\begin{equation}
\Lambda_Q=
\frac{\hbar^2}{2 m^2 c^2}~
\nabla^2 \left(
\frac{\nabla^2 {\cal R}}{{\cal R} }
\right)
=-\frac{1}{mc^2}~\nabla^2 V_Q~.
\label{lambdaq}
\end {equation}
Note that we have included a subscript $Q$ to $\Lambda$ above to emphasize that
it comes from quantum corrected boson trajectories, rather than from
the vacuum.
To estimate $\Lambda_Q$,
we go back to Eq.(\ref{cos1}) and assume a slowly varying
matter density $\rho_{crit}$.
Then Eq.(\ref{fried1}) can be written as:
\begin{eqnarray}
\ddot a + \omega^2 a=0~,
\label{ho2}
\end{eqnarray}
i.e. an harmonic oscillator with angular frequency
$\omega^2 = H_0^2/2=1/(2L_0^2)$.
Taking the harmonic potential $V$ in
Eq.(\ref{cos2}), we obtain the ground state
wavefunction
%

%Plugging in this
%potential in the Schr\"odinger equation (\ref{cos2})
%yields the following ground state
%wavefunction ${\cal R}$, with an yet
%undetermined phase $S$:
%
\begin{eqnarray}
%{\cal R}
\Psi(r,t)=R(a)~e^{-r^2/\sigma^2}e^{-iE_0t/\hbar}~.
\label{becpsi}
\end{eqnarray}
That is, ${\cal R}=R(a)~e^{-r^2/\sigma^2},~S=-E_0t$ in the above.
Also $\sigma^2=2\hbar/m\omega= 2\sqrt[]{2}\lambda L_0$,
$\lambda=\hbar/mc$ is the Compton wavelength of
the constituent bosons and $E_0=\hbar\omega/2$
is the ground state energy.
%
%we assume
%
%${\cal R}=\exp (-r^2/L_0^2)$,
%
%${\cal R}=R(t)\exp (-r^2/\sigma^2)$,
%resembling an almost flat %(and hence constant)
%wavefunction
%for the condensate, tapering off
%
%near the edge of the visible universe.
%
%at a very large distance $\sigma$.
There is an intrinsic time
dependence in the scale factor $a$, which we
assume to be slowly varying.
Furthermore, the normalization is chosen as
$R(a)=R_0/[a(t)]^{3/2}$, such that
$\rho_{crit}\propto R_0^2$,
the DM density $\rho \propto |{\cal R}|^2 \propto 1/a^3$
and BEC particle number is conserved in time.
%Such time-dependence can always be chosen
%for the (ground state) BEC wavefunction.
Note that the length scale $\sigma$ remains constant in time, and the spread of the BEC wavefunction is entirely accounted for by $R(a)$.
%Copies of this wavefunction in all directions, restores large scale homogeneity and isotropy, and likely
%contributes to the observed
%small inhomogeneities and anisotropies.
A little algebra then gives
\begin{eqnarray}
V_Q &&
%=\frac{\hbar^2}{2m \sigma^2} \left(6-\frac{4r^2}{\sigma^2}\right)
= \frac{3}{2}~\hbar\omega - \frac{1}{2} m\omega^2 r^2 \label{vqho} \\
&& = \frac{3\hbar^2}{m\sigma^2} - \frac{2\hbar^2 r^2}{m\sigma^4}
~.
\end{eqnarray}
Either comparing this with Eq.(\ref{cos1})
or from direct computation using Eq.(\ref{lambdaq}), we get
\begin{equation}
\Lambda_Q=
12 \left(\frac{\hbar}{mc} \right)^2\frac{1}{\sigma^4}
= \frac{12\lambda^2}{\sigma^4}
= \frac{3}{2 L_0^2}
\label{lambda10}
\end{equation}
That is, the small observable cosmological constant
is induced via quantum corrections.
Note that the normalization $R(a)$ or the phase
$e^{-iE_0t/\hbar}$ in Eq.(\ref{becpsi}) plays no role in above.
The above mechanism explains why it is positive
(the quantum potential is positive),
small (it is proportional to $\hbar^2$ or $1/L_0^2$, both tiny quantities) and offers a resolution of
the so-called `coincidence problem, on the approximate
equality of DM and DE - this is simply because the quantum
potential (DE) induced in a harmonic oscillator is equal and opposite to the classical harmonic potential (DM), as can be seen from
Eq.(\ref{vqho}).

As for the mass $m$,
we consider the following two cases.
First, if the constituent bosons are
gravitons, then as explained below, $m=10^{-68}$ kg or $10^{-32}eV/c^2$.
Comparing with the results of the previous section, we see that
this is well within the range of values of $m$ required for the
bosons to form a BEC and hence a DM candidate.
Next, we consider the mass preferred by fuzzy cold DM or ultralight
scalar particles
$m\approx 10^{-22}eV/c^2$, e.g.
\cite{hu,marsh,marsh2}. BEC mass of this order
is also supported by some recent work in which DM density profiles of a BEC derived from the Gross-Pitaevskii was matched with that from a number of galaxies \cite{laszlo2}.
In this case, in addition
to being in the required range for the BEC to form, this mass facilitates the formation of observed DM halos at galaxy scales.
The extent of the BEC in this case is $\sigma \approx 10^{-5} L_0$.
Bose condensation is crucial in this picture, as it is only then that
every boson has the same wave function, and the problem reduces to a
one body equation, leading to the quantum potential.

What could be these bosons in the condensate?
First, we consider massive gravitons. Although gravitons derived from general relativity
are massless, there has been considerable progress recently, both in the theoretical and
experimental fronts, in having a consistent picture of massive gravitons in extensions of
general relativity. They all in fact predict a graviton mass which is consistent with the allowed values in our work;
see for example \cite{graviton,zwicky,mann,mukhanov,oda,derham,hassan,
massive1,massive2,derham2,
hinterbichler,majid,anupam,gravmassreview}.
It was also shown recently that observations and theory impose stringent bounds on
graviton mass to the above value of $10^{-32}~eV/C^2$
\cite{alidasgrf}.
%, and that
%massive excitations of the gravitational field accompanied with a massive Higgs field can
%arise in spontaneous symmetry breaking in general relativity \cite{dasssb1,dasssb2}.
%
%
The second possibility is axions. Although axions were originally proposed
to solve the strong CP problem in
quantum chromodynamics, they also arise in the context of string theory,
and have long been advocated as DM candidates
\cite{axion1,axion2,axion3,axion4}.
BEC of axionic DM have also been explored in \cite{sikivie}.
Axion mass depends on the form of the action considered and couplings therein, but
masses in the range
${\cal O}(10^{-32}-10^{-22})~eV/c^2$ are not ruled out \cite{marsh,marsh2}.
Experiments to detect axionic DM are also in progress \cite{admx}.
It must be kept in mind however that until detected, axions remain as hypothetical
particles, requiring extension of the otherwise well-tested standard model of particle physics,
with their masses and couplings put in by hand.
Finally, as shown in \cite{dasssb1,dasssb2}, a Higgs type field in the spontaneous symmetry
breaking in general relativity is another potential candidate for the constituent bosons.

To summarize, we have shown in this article that a BEC of ultralight bosons can
account for the DM content of our universe, while its associated quantum potential
can account for DE. Our results are based on a few assumptions,
namely the existence of a BEC of light bosons spreading over cosmological distance scales and described by a slowly varying macroscopic wavefunction. The form of the wavefunction
(a Gaussian with a specific time-dependent amplitude, and repetitions thereof) ensures the correct time-dependence of DM density and that it is consistent with the observed large scale homogeneity and isotropy of  the universe.
While the main ideas here were described in an earlier paper by the authors \cite{db1}, there the issue of homogeneity and isotropy of the BEC was not examined. Furthermore, in this article we have explained the precise time-dependence of the BEC wavefunction as well as its width ($\sigma$) at any given point in time, and included the possibility of another boson as the BEC constituents, namely a gravitational-Higgs.
Our model is
robust and expected to survive even when
potential quantum corrections to spacetime itself in the very early universe
are taken into account.
The small deviations of the wavefunction from a constant,
over cosmological length scales may have observational consequences.
%
%Further, it has been claimed that bosons of our predicted mass may account for
%observed oscillations of the scale factor \cite{mead}.
We hope that this and other
testable predictions of our model will emerge in the near future.

\appendix

\section{}

Since the Friedmann equation, which the guiding equation of cosmology, can be derived
from the Raychaudhuri equation,
we start with the recently derived quantum corrected Raychaudhuri equation,
obtained by replacing geodesics with quantal (Bohmian) trajectories \cite{bohm},
associated with a wavefunction $\psi={\cal R}e^{iS}$ of the fluid or condensate filling our universe
(${\cal R} (x^\alpha), S (x^a)=$ real).
Using this, one can define the four velocity field $u_a = (\hbar/m) \partial_a S$, and expansion
$\theta=Tr(u_{a;b}) = h^{ab} u_{a;b},~h_{ab}=g_{ab}-u_au_b$, where
$a=0,1,2,3$, with $0$ signifying the time coordinate and the others the space coordinates
\cite{das1}
%
%\footnote
{(Note that
we use the metric signature $(-,+,+,+)$ here, as opposed to $(+,-,-,-)$ in \cite{das1},
resulting in opposite sign of the $\hbar^2$ terms. Here we concentrate
on the more important of the two correction terms.)}.
With this, one arrived at the relativistic Quantum Raychaudhuri equation
%
%\footnote{The Raychaudhuri equation, which is Eq.(\ref{qre2a}) without the
%${\cal O}(\hbar^2)$ terms, was derived by A. K. Raychaudhuri in 1955 \cite{akrpapers}. It is
%the key ingredient in the Hawking-Penrose singularity theorems \cite{singtheorems}.
%These theorems show that most classical spacetimes are singular and
%general relativity itself breaks down as a physical theory in -certain regions of spacetime.}
%
%
\begin{eqnarray}
\frac{d\theta}{d\lambda} =&& - \frac{1}{3}~\theta^2
- R_{cd} u^c u^d
+  \frac{\hbar^2}{m^2} h^{ab} \left( \frac{\Box {\cal R}}{\cal R} \right)_{;a;b}~,
\label{qre2a} %\\
\end{eqnarray}
where $R_{ab}$ is the Ricci tensor and $\Box \equiv \partial^a \partial_a$ is the
four dimensional d'Alembertian operator.
{The Raychaudhuri equation, which is Eq.(\ref{qre2a}) without the
${\cal O}(\hbar^2)$ terms, was derived by A. K. Raychaudhuri in 1955 \cite{akrpapers}. A similar equation
was derived around the same time by
L. D. Landau and
E. M. Lifshitz \cite{ll}.
It is
the key ingredient in the Hawking-Penrose singularity theorems \cite{singtheorems}.
These theorems show that most classical spacetimes are singular and
general relativity itself breaks down as a physical theory in -certain regions of spacetime.}
In this case too, if one ignores the last (quantum) term,
$\theta\rightarrow -\infty$ and a pencil of nearby geodesics,
known as a congruence, converge to a point at a finite proper time $\tau_0$.
The second order Friedmann equation satisfied by
the scale factor $a(t)$ can be derived from the above, by replacing \cite{akrbook}
%\bea
$\theta = 3{\dot a}/{a}~,$ %\\
%
%\theta = 2\frac{\dot a}{a}~,
%\eea
%
and
$R_{cd} u^c u^d \rightarrow \frac{4\pi G}{3} (\rho+3p) %- \Lambda c^2/3
,$
\begin{eqnarray}
\frac{\ddot a}{a} &&= - \frac{4\pi G}{3} \left( \rho + 3p \right) %+ \frac{\Lambda c^2}{3}
+ \frac{\hbar^2}{3 m^2} h^{ab} \left( \frac{\Box {\cal R}}{\cal R} \right)_{;a;b}~.
\label{frw1}
\end{eqnarray}
In this case, the quantum potential terms are the
${\cal O}(\hbar^2)$ terms in Eqs.(\ref{qre2a}) and (\ref{frw1}).
As expected, they vanish in the $\hbar\rightarrow 0$ limit, giving back
the classical Raychaudhuri and the Friedmann equations.
Note that since Bohmian trajectories do not cross
\cite{holland,nocrossing},
it follows that even when $\theta$ (or $\dot a$) $\rightarrow -\infty$, the quantal
trajectories (as opposed to classical geodesics) do not converge, and there is no counterpart of
geodesic incompleteness, or the classical singularity theorems,
and singularities such as big bang or big crunch can in fact avoided
\cite{daplb}
(the the scale factor $a$ would still be very small at the classical big bang time).
Next, we interpret the correction term as the cosmological constant
\begin{eqnarray}
\Lambda_Q = \frac{\hbar^2}{m^2 c^2} h^{ab} \left( \frac{\Box {\cal R}}{\cal R} \right)_{;a;b}~,
\label{qlambda}
\end{eqnarray}
Although $\Lambda_Q$ depends on the form of the amplitude of the
wavefunction ${\cal R}$, for any reasonable form,
such as a Gaussian wave packet in the previous section,
$\psi =
\left( R_0/[a(t)]^{3/2} \right)
\exp(-r^2/\sigma^2)$ of
extent $\sigma$,
or for one which results when an
interaction is included in the scalar field equation $[\Box + g|\psi|^2 -k ]\psi=0$, namely
$\psi=\psi_0\tanh(r/\sigma\sqrt{2}) ~(g>0)$ and $\psi=\sqrt{2}~\psi_0~{\mbox {sech}}(r/\sigma)~(g<0)$
\cite{rogel}, it can be easily shown that $(\Box {\cal R}/{\cal R})_{;a;b} \approx 1/\sigma^4$.
%where
%$L$ is the characteristic length scale in the problem, typically the
%Compton wavelength $L=\hbar/mc$ \cite{wachter}, over which the wavefunction is non-vanishing.
This again gives using $\lambda=\hbar/mc$,
\begin{eqnarray}
\Lambda_Q= 12 \left(\frac{\hbar}{mc} \right)^2\frac{1}{\sigma^4}
= 12 \frac{\lambda^2}{\sigma^4}~,
%
%\Lambda_Q = \frac{1}{L^2} =\left( \frac{m c}{\hbar} \right)^2 ~,
\label{lambda1}
\end{eqnarray}
which has the correct sign as the observed cosmological constant.
As in the previous section, the wavefunction repeats itself in space, preserving large scale homogeneity and isotropy.
Also as remarked earlier for the Newtonian derivation, this sign is expected from
the repulsive nature of quantum potential.
Furthermore, the conclusions following Eq.(\ref{lambda10}) continue to hold, and for
$\sigma = \sqrt{8L_0\lambda}$,
\begin{eqnarray}
\Lambda_Q
&& = \frac{3}{2L_0^2}~
= 10^{-52}~m^{-2} \\
&& = 10^{-123}~~(\text{in Planck units})~,
\end{eqnarray}
%
%
%Next, to estimate its magnitude,
%we note that if $L$ is identified with the linear dimension of our observable universe, $L_0$,
%$m$ can be regarded as the small mass of gravitons (or photons),
%with gravity (or Coulomb field) following a Yukawa type of force law
%
%$
%F = - \frac{Gm_1 m_2}{r^2} \exp(-r/L_0)~.
%$
%
%Since gravity (and light) has not been tested beyond the above length scale, this interpretation is
%natural, and may in fact be unavoidable.
%If one invokes periodic boundary conditions, this is also the mass of the lowest
%Kaluza-Klein modes.
%Substituting $L_0 = 1.4 \times 10^{26}~m$, one obtains
%$m \approx 10^{-68}~kg$ or $10^{-32}~eV$,
%quite consistent with the estimated bounds on graviton masses from various experiments, as noted in the previous section.
%As observed in the previous section, gravitons, axions and the gravitational Higgs field can all
%be viable candidates for the bosons making up the BEC, and further research is required to
%lend support or rule out one or more of them.
% In \cite{derham} too, $\Lambda \sim m^2$ was obtained from field theoretic considerations.
%
%Finally, plugging in the above value of $L_0$ in Eq.(\ref{lambda1}), we get
%
%\begin{eqnarray}
%\Lambda_Q && = 10^{-52} m^{-2} \\
%%
%&& = 10^{-123}~~(\text{in Planck units})~,
%\end{eqnarray}
%
since the Planck length
$\ell_{Pl} = 1.6 \times 10^{-35}$ m.
%The above is indeed the observed value of $\Lambda$.
%Also, since the size of the observable universe is about $c/H_0$, where
%$H_0$ is the current value of the Hubble parameter, one sees why
%the above value of $\Lambda_Q$ numerically equals $H_0^2/c^2$
%(which is $8\pi G/3 c^4 \times \rho_{crit}$, the critical density), offering a viable
%explanation of the coincidence problem.

In summary, results identical to the ones obtained in the previous sections are
reproduced in this Appendix, although from a more rigorous and completely relativistic
point of view. This in turn justified the use of Newtonian cosmology to arrive at our
main results.

\vspace{.2cm}
%%%%%%%%%%%%%%%%%%%%%%%%%%%%%%%%%%%%%%%%%%%%%%%%%%%%%%
\noindent{\bf Acknowledgment}

\noindent
% This article is dedicated to the memory of Professor A. K. Raychaudhuri, respected teacher
% of one of the authors (SD).
%
This work was supported by the Natural Sciences and Engineering
Research Council of Canada.
%%%%%%%%%%%%%%%%%%%%%%%%%%%%%%%%%%%%%%%%%%%%%%%%%%%%%


\begin{thebibliography}{99}

\bibitem{mond1}
M. Milgrom,
New  Astron. Rev. {\bf 46}, 741 (2002).

\bibitem{mond2}
S. S. McGaugh,
Phys. Rev. Lett. {\bf 106}, 121303 (2011).

\bibitem{altdm}
J. D. Bekenstein,
From 'Particle Dark Matter: Observations, Models and Searches', edited by G. Bertone (Cambridge U. Press, Cambridge 2010) Chap.6, p.95-114
[arXiv:1001.3876].

\bibitem{altde}
M. Sami, arXiv:0901.0756.

\bibitem{dereview1}
E. J. Copeland, M. Sami. S. Tsujikawa, Int. J. Mod. Phys.
{\bf D15}, 1753-1936 (2006)
[arXiv:hep-th/0603057].

\bibitem{dereview2}
J. Frieman, M. Turner, D. Huterer
Ann. Rev. Astron. Astrophys.
{\bf 46},385-432 (2008)
[arXiv:0803.0982].


\bibitem{dmreview1}
B-L. Young, Front. Phys.12: 121201 (2017).

\bibitem{dmreview2}
T. Plehn, arXiv:1705.01987.

\bibitem{bose}
S. N. Bose, Zeit. f\"ur Physik,
{\bf 26}, 178 (1924).

\bibitem{einstein}
A. Einstein, Sitz. der Preussischen Acad.
der Wissen. {\bf 1}, 3 (1925).

\bibitem{db1}
S. Das, R. K. Bhaduri,
Class. Quant. Grav. {\bf 32}, 105003 (2015) [arXiv:1411.0753].

\bibitem{Hu2000} W. Hu{\it et. al} {Phys .Rev. Lett. 2000 85 1158 }

\bibitem{Lopez}L. A. Urena-Lopez, J of Cosmology and Astroparticle Phys. 01 2009 14
Arxiv 0806.3093.

\bibitem{Bohua} Bohua Li, M.A. Thesis, Univ. Texas (2013).

\bibitem{sudarshan}
K. P. Sinha, C. Sivaram, E. C. G. Sudarshan,
Found. Phys. {\bf 6}, no.1 (1976) 65;
Found. Phys. {\bf 6}, no.6 (1976) 717.

\bibitem{khlo1}
M. Yu. Khlopov, B. A. Malomed and Ya. B. Zeldovich, Mon. Not. Roy. astr. Soc. {\bf 215}, 575-589 (1985).

\bibitem{khlo2}
M. Yu. Khlopov, A. S. Sakharov and D. D. Sokoloff,
Nucl.Phys. {\bf B} (Proc. Suppl.) {\bf 72}, 105-109 (1999).

\bibitem{morikawa} M. Morikawa,
22nd Texas Symp. on Rel. Astro. (2004) 1122;
T. Fukuyama, M. Morikawa, Prog. Theo. Phys. {\bf 115}, no. 6 (2006) 1047-1068.

\bibitem{moffat} J. W. Moffat, astro-ph/0602607.

\bibitem{wang} X. Z. Wang, Phys. Rev {\bf D64} (2001) 124009.

\bibitem{boehmer} C. G. Boehmer, T. Harko, JCAP 0706:025,2007 [arXiv:0705.4158];
T. Harko, G. Mocanu, Phys. Rev. {\bf D85} (2012) 084012 [arXiv:1203.2984].

\bibitem{sikivie} P. Sikivie, arXiv:0909.0949.

\bibitem{chavanis} P-H. Chavanis, A \& A {\bf 537} (2012) A127 [arXiv:1103.2698].

\bibitem{dvali} G. Dvali, C. Gomez, Fortsch. Phys. {\bf 61} (2013) 742-767 [arXiv:1112.3359].

\bibitem{houri}
H. Ziaeeapour, arXiv:1112.3934.

\bibitem{kain} B. Kain, H. Y. Ling, Phys. Rev. {\bf D85} (2012) 023527 [arXiv:1112.4169].

\bibitem{suarez} A. Su\'arez, V. Robles, T. Matos,
Astroph. and Space Sc. Proc. {\bf 38} Chapter 9 (2013) [arXiv:1302.0903].

\bibitem{ebadi} Z. Ebadi, B. Mirza, H. Mohammadzadeh, JCAP 11(2013)057 [arXiv:1312.0176].

\bibitem{laszlo1} M. Dwornik, Z. Keresztes, L. A. Gergely,
{\it Recent Development in Dark Matter Research},
Eds. N. Kinjo, A. Nakajima, Nova Science Publishers (2014), p.195-219 [arXiv:1312.3715];
arXiv:1406.0388.

\bibitem{bettoni} D. Bettoni, M. Colombo, S. Liberati,
JCAP02(2014)004 [arXiv:1310.3753].

\bibitem{gielen} S. Gielen, arXiv:1404.2944.

\bibitem{schive} H. Schive, T. Chiueh, T. Broadhurst, Nature Physics {\bf 10} (2014) 496
[arXiv:1406.6586].

\bibitem{davidson} S. Davidson, Astrop. Phys.{\bf 65} (2015) 101 [arXiv:1401.1139].

\bibitem{casadio1}
M. Cadoni, R. Casadio, A. Giusti, W. M\"uck, M. Tuveri,
Phys. Lett. {\bf B776},
(2018) 242 (2018) [arXiv:1707.09945].
\bibitem{casadio2}
M. Cadoni, R. Casadio, A. Giusti, M. Tuveri,
Phys. Rev. {\bf D97}, 044047 (2018)
[arXiv:1801.10374].


%\bibitem{Arnab} {Arnab Rai Choudhuri {\it Astrophysics for Physicists} , Cambridge University Press, 2010, page 308}

\bibitem{weinberg} S. Weinberg, {\it Cosmology}, Cambridge.

\bibitem{perlmutter} S. Perlmutter et al.
Astrophysical J. {\bf 517 (2)} (1999) 565�86 [arXiv:astro-ph/9812133].

\bibitem{riess} A. G. Riess et al., Astron. J. {\bf 116} (1998) 1009 [arXiv:astro-ph/9805201].

\bibitem{Rindler} {Wolfgang Rindler {\it Essential Relativity} Springer Verlag, Second Edition  1979 , pp.224-225.}


\bibitem{inflation}
D. Baumann, TASI Lectures on Inflation,
arXiv:0907.5424.

\bibitem{brack} M. Brack, R. K. Bhaduri, {\it Semiclassical Physics}, Westview Press (2003).

\bibitem{grether}
M.Grether, M.de Llano and G.A.Baker Jr. Phys.Rev.
Lett. {\bf 99} 200406 (2007).

\bibitem{fujita} S. Fujita, T. Kimura, Y. Zheng,
Found. Phys. {\bf 21}, no.9 (1991) 1117.

\bibitem{bhaduricjp} W. van Dijk, C. Lobo, A. MacDonald, R. K. Bhaduri,
Can. J. Phys. {\bf 93} (8), 830-835 (2015).
%arXiv:1412.5112 (to appear in Can. J. Phys.).

\bibitem{Haber} {H. E. Haber and H. A. Weldon, Phys.Rev.Lett. {\bf 46}, 1497 (1981)}

\bibitem{landau}
L. D. Landau and E. M. Lifshitz,
{Statistical Physics, 3rd Edition, Part 1},
Pergamon (1959).

%\bibitem{sami} E. J. Copeland, M. Sami. S. Tsujikawa, Int.J.Mod.Phys.D15:1753-1936 (2006)
%[arXiv:hep-th/0603057].


\bibitem{bohm}
D. Bohm, Phys. Rev. {\bf 85} (1952) 166; D. Bohm, B. J.
 Hiley, P. N. Kaloyerou, Phys. Rep. {\bf 144}, No.6 (1987) 321.

\bibitem{holland}
P. R. Holland, {\it The Quantum Theory of Motion}, Cambridge (1993).

\bibitem{akrbook}
A. K. Raychaudhuri, {\it Theoretical Cosmology}, Oxford (1979).

\bibitem{das1}
S. Das, Phys.Rev.{\bf D89} 084068 (2014).

\bibitem{das2}
S. Das, Int. J. Mod. Phys. {\bf 24} 1442017 (2014).

\bibitem{hu}
W. Hu, R. Barkana, A. Gruzinov,
Phys. Rev. Lett. {\bf 85}, 1158-1161 (2000) [arXiv:astro-ph/0003365].

\bibitem{marsh} R. Hlozek, D. Grin, D. J. E. Marsh, P. G. Ferreira, arXiv:1410.2896.

\bibitem{marsh2} D. Marsh, A. Pop,
MNRAS, 451, 2479 (2015).
[arXiv:1502.03456].

\bibitem{laszlo2}
E. Kun, Z. Keresztes, S. Das,
L. A. Gergely (in preparation).

\bibitem{graviton}
C. M. Will, Phys. Rev. {\bf D57} (1998) 2061;
L. S. Finn, P. J. Sutton Phys. Rev. {\bf D65} (2002) 044022 [arXiv:gr-qc/0109049];
A. S. Goldhaber, M. M. Nieto, Rev. Mod. Phys. {\bf 82} (2010) 939 [arXiv:0809.1003]
E. Berti, J, Gair, A. Sesana, Phys. Rev. {\bf D84} (2011) 101501(R).

\bibitem{zwicky} F. Zwicky,
{\it Cosmic and terrestrial tests for the rest mass of gravitons},
Publications of the Astronomical Society of the Pacific, Vol. 73, No. 434, p.314.

\bibitem{mann} J. R. Mureika, R. B. Mann,
Mod. Phys. Lett. {\bf A26}  (2011) 171-181 [arXiv:1005.2214].

\bibitem{mukhanov} A. H. Chamseddine, V. Mukhanov, JHEP08(2010) 011 [arXiv:1002.3877].

\bibitem{oda} I. Oda, arXiv:1003.1437.

\bibitem{derham} C. de Rham, G. Gabadadze, L. Heisenberg, D. Pirtskhalava,
Phys. Rev. {\bf D83} (2011) 103516 [arXiv:1010.1780];
C. de Rham, G. Gabadadze, A. J. Tolley,
Phys. Rev. Lett. {\bf 106} (2011) 231101 [arXiv:1011.1232];

\bibitem{hassan} S. F. Hassan, R. A. Rosen, Phys. Rev. Lett. {\bf 108} (2012) 041101 [arXiv:1106.3344].

\bibitem{massive1}
G. Leon, J. Saavedra, E. N. Saridakis, Class. Quant. Grav. {\bf 30} (2013) 135001 [arXiv:1301.7419].

\bibitem{massive2}
R. Gannouji, M. W. Hossain, M. Sami, E. N. Saridakis,
Phys. Rev. {\bf D87} (2013) 123536 [arXiv:1304.5095].

\bibitem{derham2} C. de Rham, M. Fasiello, A. J. Tolley, Int. J. Mod. Phys. {\bf D23}, no. 13, 14430006 (2014) [arXiv:1410.0960].

\bibitem{majid} S. Majid, arXiv:1401.0673.

\bibitem{hinterbichler} K. Hinterbichler,
Rev. Mod. Phys. {\bf 84} (2012) 671-710 [arXiv:1105.3735].

\bibitem{anupam} A. Ashoorioon, P. S. Bhupal Dev, A. Mazumdar,
Mod. Phys. Lett. {\bf A29} (2014) 1450163 [arXiv:1211.4678].

\bibitem{gravmassreview}
C. de Rham, J. Tate Deskins. A. J. Tolley, S-Y Zhou,
Rev. Mod. Phys. {\bf 89}, 025004 (2017).


\bibitem{alidasgrf}
A. F. Ali, S. Das,
Int. J. Mod. Phys. {\bf 25} 1644001 (2016).

\bibitem{axion1}
L.F. Abbott, P. Sikivie, Phys. Lett. B 120, 133 (1983).

\bibitem{axion2}
J. Preskill, M.B. Wise, F. Wilczek, Phys. Lett. B 120,
127 (1983).

\bibitem{axion3}
M. Dine, W. Fischler,
Phys.Lett. {\bf B120} 137 (1983).

\bibitem{axion4}
M.S. Turner, Phys. Rev. D 28, 1243 (1983).

\bibitem{admx} S. Aztalos et al (ADMX collaboration), Phys. Rev. Lett. {\bf 104}
(2010) 041301 [arXiv:0910.5914].

\bibitem{dasssb1}
S. Das, M. Faizal, E. C. Vagenas,
Int. J. Mod. Phys. D (to appear),
arXiv: 1805.05665.

\bibitem{dasssb2}
S. Das, M. Faizal,
Gen. Rel. Grav. {\bf 50}, 87 (2018).

\bibitem{akrpapers}
A. K. Raychaudhuri, Phys. Rev. {\bf 98}, 1123 (1955).

\bibitem{ll}
L. D. Landau, E. M. Lifshitz,
Classical Theory of Fields
(1951).

\bibitem{singtheorems}
R. Penrose, Phys. Rev. Lett. {\bf 14},
57 (1965);
S. W. Hawking and R. Penrose, Proc. R. Soc. {\bf A314},
529 (1970).


\bibitem{nocrossing}
C. Phillipidis, C. Dewdney, B. J. Hiley, Nuovo Cimento,
{\bf 52B} (1979) 15-28;
D. A. Deckert, D. D\"urr, P. Pickl, J. Phys. Chem. A, 111,
41 (2007) 10325; A. S. Sanz, J. Phys.: Conf. Ser. {\bf 361},
012016 (2012);
%
A. Figalli, C. Klein, P. Markowich, C. Sparber,
{\it WKB analysis of Bohmian dynamics}, arXiv:1202.3134.


%\bibitem{ahmed} A. Awad, A. F. Ali, B. Majumder, JCAP {\bf 10} (2013) 052 [arXiv:1308.4343].

\bibitem{daplb}
A. F. Ali, S. Das,
Phys. Lett. {\bf B741}, 276 (2015)
[arXiv:1404.3093].

\bibitem{rogel} J. Rogel-Salazar, Eur. J. Phys. {\bf 34} (2013) 247 [arXiv:1301.2073];
N. \"Uzar, S. Deniz Han, T. T\"ufekci, E. Aydiner, arXiv: 1203.3352.

%\bibitem{wachter} A. Wachter, {\it Relativistic Quantum Mechanics}, Springer (2010).


%\bibitem{mead}

\end{thebibliography}
\end{document}